\documentclass[aps,prl,letterpaper,twocolumn,showpacs,superscriptaddress,floatfix,longbibliography]{revtex4-1}

\usepackage[english]{babel}
\usepackage{amsmath}
\usepackage{color}
\usepackage{epsfig}
\usepackage{graphicx}
\usepackage{bm}
\usepackage{times,amsmath,amssymb}
\usepackage{amsfonts}
\usepackage{amssymb}
\usepackage{amsmath}
\usepackage{color}
\usepackage{mathtools}
\usepackage{empheq}

\global\long\def\bra#1{\langle #1 |}
\global\long\def\ket#1{| #1 \rangle }

\global\long\def\al{\alpha}
\global\long\def\no{\nonumber}
\global\long\def\be{\beta}
\global\long\def\ga{\gamma}

\global\long\def\De{\Delta}

\global\long\def\th{\theta}

\global\long\def\en{E}
\global\long\def\la{\lambda}
\global\long\def\ka{\kappa}
\global\long\def\si{\sigma}
\global\long\def\vfi{\gamma}

\newcommand{\nc}{\newcommand}
\nc{\ir}{\mathrm{i}}
\nc{\dd}{\mathrm{d}}   
\nc{\eE}{\mathsf{e}}
\nc{\mom}{k}
\nc{\kind}{l}

\begin{document}

\title{Phantom Bethe excitations  and spin helix eigenstates in integrable periodic and open spin  chains
}
\author{Vladislav Popkov}
 \affiliation{Department of Physics,
  University of Wuppertal, Gaussstra\ss e 20, 42119 Wuppertal,
  Germany}
 \affiliation{Faculty of Mathematics and Physics, University of Ljubljana, Jadranska 19, SI-1000 Ljubljana, Slovenia}
\author{Xin  Zhang}
 \affiliation{Department of Physics,
  University of Wuppertal, Gaussstra\ss e 20, 42119 Wuppertal,
  Germany}
\author{  Andreas Kl\"umper} 
 \affiliation{Department of Physics,
  University of Wuppertal, Gaussstra\ss e 20, 42119 Wuppertal,
  Germany}

\begin{abstract}

We demonstrate the existence of a special chiral ``phantom" mode with some
analogy to a Goldstone mode in the anisotropic quantum $XXZ$ Heisenberg spin
chain.  The phantom excitations contribute zero energy to the eigenstate, but a
finite fixed quantum of momentum $\mom_0$. The mode exists not due to symmetry
principles, but results from non-trivial scattering properties of magnons with
momentum $\mom_0$ given by the anisotropy via $\cos \mom_0=\De$.  Different
occupations of the phantom mode lead to energetical degeneracies between
different magnetization sectors in the periodic case.  This mode
originates from special string-type solutions of the Bethe ansatz equations
with unbounded rapidities, the phantom Bethe roots (PBR).  We derive
criteria under which the spectrum contains eigenstates with
PBR, both in open and periodically closed integrable systems, for spin $1/2$ and higher
spins, and discuss the respective chiral eigenstates.  The simplest of such
eigenstates, the spin helix state which is a periodically modulated state of chiral
nature, is built up from the phantom excitations exclusively. Implications of
our results for experiments are discussed.

\end{abstract}

\maketitle



Interacting quantum spin systems are a vibrant research field as fascinating
kinds of order are realized with rather complex order parameters or of
topological nature.  Even the spin-1/2 $XXZ$ chain, despite its long history
and being one of the best studied paradigmatic models in quantum statistical
mechanics \cite{GaudinBook}, remains a source of inspiration and fascinating
new progress. This model is integrable and in principle allows for the
calculation of objects that in generic systems are usually not accessible in
the thermodynamic limit. Among the relatively
recent results the discovery of a set of quasi-local conserved quantities
\cite{2013ProsenQuasilocal} with strong implications on the theory of
finite-temperature quantum transport \cite{2020BertiniReview} and successes in
the calculation of finite temperature correlation functions
\cite{GoehmannKluemperSeel,BoosGoehmann} are exciting achievements.

In this letter we are interested in phenomena of anisotropic quantum spin
chains requiring the understanding of energetical degeneracies in uncharted
territory. A first example is the physics of so-called spin helix states (SHS)
(\ref{eq:SHS}) which show sharp local polarization with respect to site
dependent axes. These states are routinely created, and widely used in
coherent experimental protocols \cite{2020NatureSpinHelix,2021Ketterle,
  2014HildSHS}.  SHS can also be generated as non-equilibrium steady states
via a dissipative quantum protocol
\cite{2016PopkovPresilla,2017PopkovSchutzHelix,2020ZenoPRL} or via controlled
local boundary dissipation.  Remarkably, the needed boundary dissipation is of
the type which allows the system to retain, partly, its integrability
\cite{ProsenReview2015}.

The eigenvalue degeneracies of isotropic quantum spin chains are well
understood on the basis of the $su(2)$ symmetry algebra. 
Simple eigenstates, that are fully polarized  with respect to any axis form a multiplet of 
degeneracy $N+1$ for the spin-1/2 chain of length $N$.

The high degeneracy of this ferromagnetic multiplet can alternatively be
explained by magnon excitations with soft Goldstone mode at wavenumber
$\mom=0$. Contrary to the usual situation when all magnons carry different
momenta, this precise $\mom=0$ mode can be occupied up to $N$ times.

A $z$-anisotropy of the spin exchange interaction lifts the high degeneracy,
leaving just two degenerate eigenstates with spins fully polarized in $+z$ or
$- z$ direction.  The $su(2)$-type degeneracies can be restored by the
so-called ``quantum deformation'' $U_q(su(2))$ of the symmetry algebra
\cite{Drinfeld,Jimbo,Jimbo1986q}, involving special possibly non-hermitian
boundary terms.

 Remarkably, an analogue of a Goldstone mode scenario can happen in periodic
 spin systems with $z$-exchange anisotropy, namely a multiple occupation of a
 single mode can occur, but now with a nonzero wave vector $\mom_0$ fine-tuned
 to the system's anisotropy $J_z/J_x= \De$ via $\cos \mom_0 = \De$.  The
 corresponding excitation can be created at zero energetic cost.  Like in the
 isotropic case, the possibility of multiple occupations of the same
 zero-energy phantom mode leads to the high degeneracy. Unlike in the
 isotropic case, the eigenstates form a multiplet of degenerate chiral states
 carrying finite current.

Excitations with momentum mode $\pm k_0$ were discussed in
  \cite{Ref1,Ref2,Ref3,Ref4,Ref5,Ref6} for accounting for the energetical
  degeneracies of the spin-1/2 XXZ chain and related systems. For certain
  systems with commensurable values of $k_0$ extended symmetry algebras are
  realized and the completeness of the Bethe ansatz has been investigated
  \cite{Ref2,Ref3,Ref4,Ref5,Ref6}.


 In our letter we show why a macroscopic occupation of precisely $\pm \mom_0$
 becomes possible, despite magnons of the wave number $\mom_0$ have
 non-trivial scattering. These states are realized by non-standard string-type
 solutions of the Bethe ansatz equations with infinite rapidities. The Bethe
 Ansatz equations for singular roots are satisfied with a universal choice for
 their arrangement (\ref{Res:PhantomBetheRoots}), which makes them effectively
 ``disappear'' from the set of Bethe Ansatz equations.
For this reason we call the roots with infinite rapidities phantom Bethe roots
and the respective excitations phantom excitations.

We find
phantom Bethe roots in
other integrable systems including open quantum systems and also 
for higher spins.

Finally, we find that the     role of the fully polarized  eigenstates in the isotropic case is taken,
in anisotropic systems,  by  simple but rather nontrivial chiral states, the spin helix states.
The SHS have ballistic current and  a harmonic modulation 
 (with period $2\pi/\mom_0$) of  transversal magnetization. 
Remarkably,  the SHS  are created with  exclusively
phantom excitations, both in open and in periodic spin chains.

\textit{Factorized eigenstates at
  commensurate values of anisotropy.}  We consider the $XXZ$ spin-1/2
Hamiltonian for periodic and open boundary conditions.  For the periodically
closed chain we have
\begin{align}
 & H_{XXZ} =\sum_{n=1}^{N}  h_{n,n+1}(\De)\,,\label{eq:XXZ}\\
& h_{n,n+1}(\De)=J \left[\sigma^x_n \sigma_{n+1}^x
  +  \sigma^y_n \sigma_{n+1}^y+\Delta\left(\sigma^z_n \sigma_{n+1}^z -
  I\right)\right],\nonumber
\end{align}
with boundary conditions $\vec\sigma_{N+1}\equiv \vec\sigma_1$.  For
  convenience we put $J=1$ throughout this letter.
For the open chain
we have
\begin{align}
H_{XXZ}&= \sum_{n=1}^{N-1}  h_{n,n+1}(\De) + \vec{h}_l \vec{\sigma}_{1} +  \vec{h}_r \vec{\sigma}_{N}\,,
\label{eq:XXZopen}
\end{align}
with boundary fields $\vec{h}_l$ and $\vec{h}_r$ on the first and on the last
sites. In both cases a shift $- J \De$ in the nearest-neighbour interaction
(\ref{eq:XXZ}) is added for convenience. Both models (\ref{eq:XXZ}),
(\ref{eq:XXZopen}) are integrable and solvable via Bethe Ansatz methods
\cite{Baxter,FaddeevTakhtajan,SklyaninFaddeevTakhtajan}.  We parametrize the
anisotropy $\De$ of the exchange interaction as $\De= \cos \gamma$ or $\De=
\cosh \eta$ with $\eta=\ir\gamma$.

We want to construct factorized eigenstates of the Hamiltonians and
introduce for each site the qubit state 
\begin{align}
&\ket{y}=\binom {1}{ \eE^y}\,. \label{DefQubit}
\end{align}
The qubit state (\ref{DefQubit})  with $y=f + \ir F$ corresponds to a fully polarized spin $1/2$
pointing into the direction $\vec{n}=(\sin \th\cos F, \sin \th\sin F,\cos \th)$
with $\tan \frac{\th}{2}= \eE^f$. 
A site-factorized state, the so-called spin-helix state (SHS)
\cite{2016PopkovPresilla,2017PopkovSchutzHelix}
\begin{align}
\ket{{SHS}(y_0,\varphi)} &=  \ket{y_0}_1  \ket{y_0+ \ir \varphi}_2 \ldots  \ket{y_0+ \ir(N-1) \varphi}_{N},
  \label{eq:SHS}
\end{align}
subscripts indicating the site number, with uniformly increasing angles
  on some offset $y_0$ becomes an eigenstate of the $XXZ$ Hamiltonian if: (i)
  the increase $\varphi$ of the angle is identical to $\pm \gamma$, the
  parameter of the anisotropy $\De=\cos \vfi$, and (ii) the boundary
  conditions can be accounted for. The parameter $\varphi$ is real (imaginary)
  for easy plane (easy axis) anisotropy corresponding to a state with
  uniformly increasing azimuthal (polar) angle.

The bulk interaction of the $XXZ$ Hamiltonian applied to any SHS state
(\ref{eq:SHS}) results in 0 due to the 
``divergence'' relation
\begin{align}
  &h(\De) \ket{y}\otimes \ket{y+\ir \vfi} =\no\\
&\ = \ket{y}\otimes (\ka\si^z \ket{y+\ir
  \vfi})- (\ka\si^z \ket{y}) \otimes \ket{y+\ir \vfi}\,,
\label{eq:Divergence}
\end{align}
where $\ka=  \ir \sin \vfi$.  For the periodic model (\ref{eq:XXZ}), the SHS
  will be an eigenstate if the periodic closure condition $\vfi N=
2\pi m$ with integer $m$ is satisfied.
This can only happen for anisotropy $|\De|\le1$.  

\begin{figure}[tbp]
\centerline{
\includegraphics[width=0.46\textwidth]{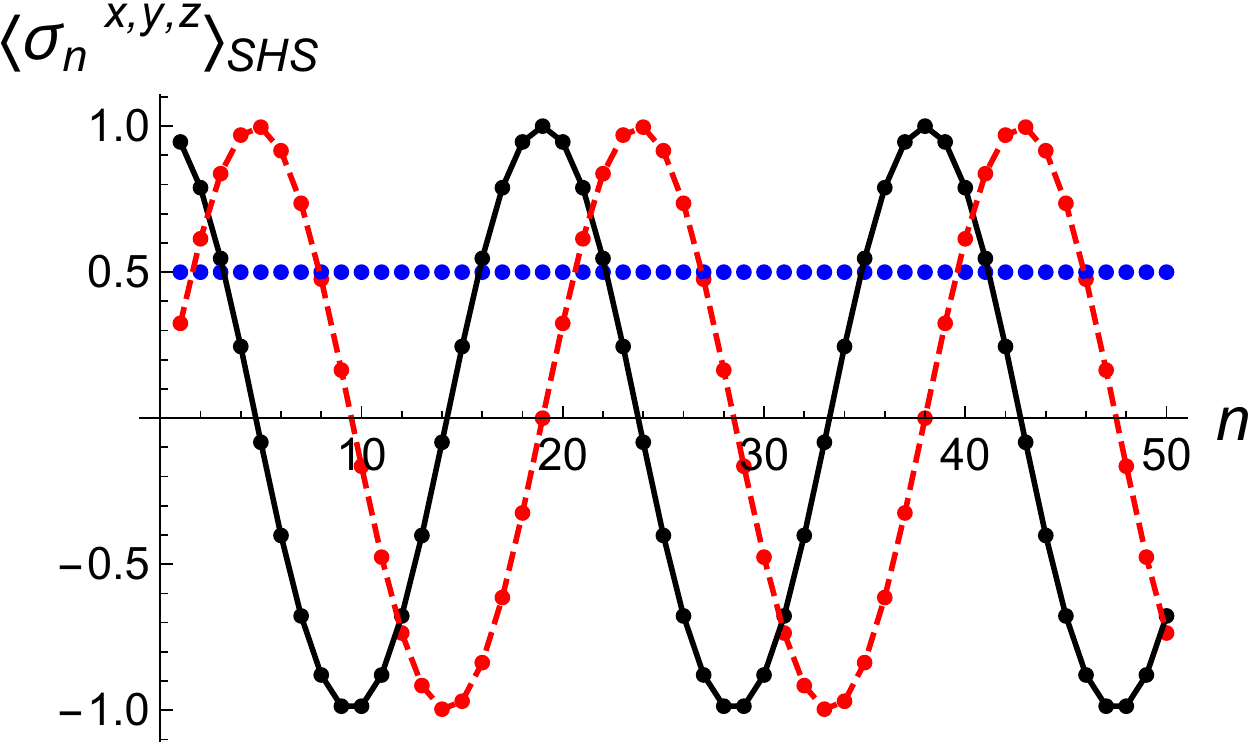}}\vspace{5mm}
\centerline
{\includegraphics[width=0.46\textwidth]{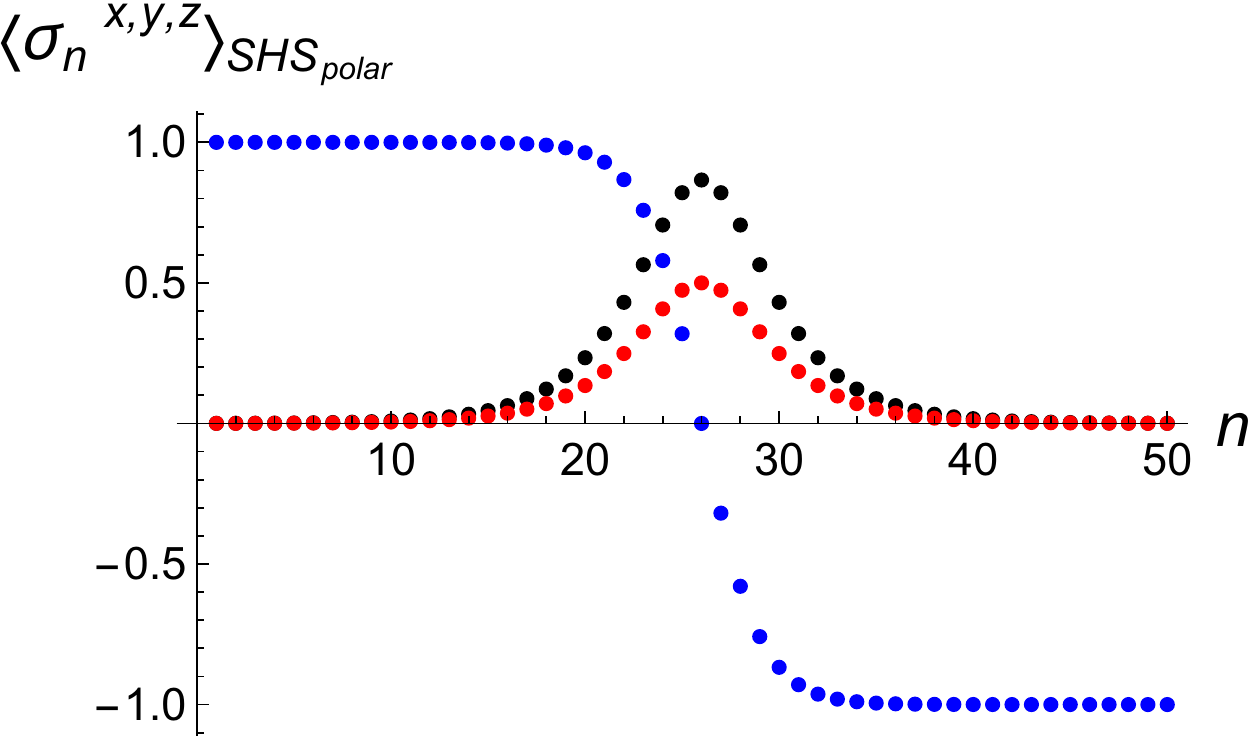}
}
\caption{Components of local magnetization $\langle \si_n^x \rangle,\langle
  \si_n^y \rangle, \langle \si_n^z \rangle$ for SHS/phantom Bethe states
  versus site number $n$, for the easy plane (upper panel) and the easy axis
  case (lower panel), indicated with black, red and blue points respectively.
  \textbf{Upper panel:} SHS (\ref{eq:SHS}) with increasing azimuthal angle,
  the phantom Bethe eigenstate of (\ref{eq:XXZopen}) or (\ref{eq:XXZ}) for
  $|\De|<1$.  Parameters: $\varphi=\vfi=2\pi/19$, $y_0= \ir \vfi +
  1/\sqrt{3}$. Curves connecting points serve as a guide for the eye.
  \textbf{Lower panel:} SHS (\ref{eq:SHS}) with increasing polar angle, the
  phantom Bethe eigenstate of (\ref{eq:XXZopen}) for $\De>1$.  Parameters:
  $\ir\varphi=\eta=2 \pi/19, y_0= \ir \pi/6+N\eta/2$.  }
\label{Fig-SHSCosCosh}
\end{figure}

For the open chain condition (ii) on the boundary
can be satisfied not only in the case $|\De|\le1$, but also for $|\De|>1$. For
$|\De|>1$ we may use expression (\ref{eq:SHS}) with the replacement
$\varphi=\ir\eta$ which results in a spin-helix state with fixed azimuthal
angle and uniformly increasing polar angles.  The
eigenstate condition is fulfilled, if the boundary interactions
$h_{l}=\vec{h}_{l} \vec{\sigma}_{1}$ and $h_{r}=\vec{h}_{r} \vec{\sigma}_{N}$
satisfy
\begin{align}
h_l \ket{y_0}&= \ka \,\si^z \ket{y_0} + \la_{-}
\ket{y_0}\,, \label{eq:hlAction}\\
h_r \ket{y_{N-1}}&= -\ka\,
  \si^z\ket{y_{N-1}}
  + \la_{+}\,
  \ket{y_{N-1}}\,, \label{eq:hrAction}
\end{align}
where $y_{N-1}=y_0+\ir (N-1)\vfi$, and $\la_\pm$ are some
  boundary-dependent constants.  The energy eigenvalue
is $E=\la_{-}+\la_{+}$.  Note that in the open chain case a condition on
the anisotropy $\De$ like in the periodic case is absent and $\varphi$ in
(\ref{eq:SHS}) can be real or imaginary.
Although having the same algebraic form, the SHS for the easy plane and
  easy axis cases have rather different physical properties as
  visualized in Fig.~\ref{Fig-SHSCosCosh}.

The factorized SHS state is after the ferromagnetic state the simplest
eigenstate of $XXZ$ spin chains.  Yet the SHS (\ref{eq:SHS}) is quite nontrivial,
and describes a ``frozen'' spin precession around the $z$-axis with period
$2\pi/\varphi$, see Fig.~\ref{Fig-SHSCosCosh}.
Due to the chiral nature, the SHS carries a
remarkably high magnetization current, finite
in the thermodynamic limit:
\begin{align*}
 \langle j^z\rangle_{SHS} &=\langle 4 \ir ( \si_n^{+}\si_{n+1}^{-} - h.c. )\rangle_{SHS}
=  \pm 2 \frac{\sin \vfi}{ \,\cosh^2 ({\rm Re}[y_0])},
\end{align*}
where the sign $\pm$ corresponds to the choice $\varphi = \pm \gamma$ in
(\ref{eq:SHS}). Remarkably, the SHS (\ref{eq:SHS}) with adjustable
  wavelength can be realized in cold atom experiments
  \cite{2020NatureSpinHelix,2021Ketterle}.  

The very existence of an eigenstate (\ref{eq:SHS}) for the periodic spin chain,
characterized by periodic modulations in the magnetization profile seems to
contradict the $U(1)$ symmetry: $XXZ$ eigenvectors
split in blocks with well defined values of the global magnetization $S^z= \sum_n 
\si_n^z$ and expectation values $ \langle \si_n^{+} \rangle =\langle \si_n^{-} \rangle
=0$ vanish, and so do $ \langle \si_n^{x} \rangle=\langle
\si_n^{y} \rangle=0$.

This paradox is resolved by the energetical degeneracy of eigenstates with
different values of the total magnetization $S^z$. We will show that  a superposition of states from
different blocks yields the state (\ref{eq:SHS}) which is not an eigenstate of
the operator $ {S}^z$.


\textit{Phantom Bethe roots at  commensurate anisotropies in periodic $XXZ$ chains.}
The eigenstates and eigenvalues are given in terms of rapidities $\mu_j$
($j=1,2,\ldots n$) whose total number $n$ may take any value out of $0,1,\ldots N$.
For any solution of the Bethe Ansatz equations (BAE)
\begin{align}
&\frac{\sinh^N (\mu_j-\ir \vfi/2)}{\sinh^N (\mu_j+\ir \vfi/2)} =\prod_{\kind\neq j}^n
\frac{\sinh (\mu_j-\mu_\kind-\ir \vfi)}{\sinh (\mu_j-\mu_\kind+\ir \vfi)},\label{eq:BAE}
\end{align}
there is an eigenstate with energy and total momentum
\begin{align}
  &\en=-\sum_{j=1}^n e(\mu_j),
  \quad K=\sum_{j=1}^n \mom(\mu_j)\,,
\label{eq:BetheEigenvalue}
\end{align}
with  single particle energy and momentum defined by
\begin{align}
 e(\mu_j)=\frac{4 \sin^2 \vfi }{ \cosh(2\mu_j)-\cos\vfi}\,,  \ \   \eE^{\ir \mom(\mu) } =\frac{\sinh
   (\mu+ \ir\frac{\vfi}{2})}{\sinh (\mu-\ir \frac{\vfi}{2})}.
 \label{singlspartmoment}
\end{align}
The Bethe eigenvector $\Psi_{\mu_1,\ldots \mu_n}=B(\mu_1)\ldots B( \mu_n) \ket{0}$
is obtained by the application of magnon creation operators $B(\mu_j)$ to the 
reference state $\ket{0}=\ket{\uparrow \uparrow \ldots \uparrow }$ 
of fully polarized spins \cite{FaddeevTakhtajan,SM}. 

\textit{Definition.} We shall call a Bethe root $\mu_p$ satisfying
(\ref{eq:BAE}), a \textit{phantom} Bethe root, if it does not give a
contribution to the respective energy eigenvalue (\ref{eq:BetheEigenvalue})
i.e.~if
${\rm Re}[\mu_p]=  \pm \infty$.
The next Lemma affirms that such phantom Bethe roots do exist: \textbf{ Lemma
  1:} For anisotropy $\vfi=2\pi m/N$ with integer $m$ there exist the
following ``phantom" solutions of the BAE (\ref{eq:BAE}) for any given
$n=1,2,\ldots N$
\begin{align}
&\mu_p=\pm \infty + \ir \pi \frac{p}{n}, \quad p=1,2\ldots n. \label{Res:PhantomBetheRoots}
\end{align}
These distributions remind of the string solutions to the Bethe ansatz
equations. Note however that (\ref{Res:PhantomBetheRoots}) holds for any
finite system size $N$ with a total number $n$ of roots 
equidistantly distributed with separation $\pi/n$. 
Note that our lemma describes the precise
arrangement of the infinite roots appearing in [1-6].  Upon
  introducing a finite magnetic flux resp.~twisted boundary conditions, the
  roots become finite while the imaginary parts stay
  close to the values of Lemma 1. This is relevant for the dependence of the
  energy as function of the twist and has important consequences for the transport
  properties, see \cite{SM}.

\textbf{Proof.} Assume $\mu_j= \pm \mu_\infty + \ir \pi {j}/{n}$, where
$\mu_\infty$ has a large real part which we let to $\infty$ when evaluating the
LHS of the Bethe ansatz equations. 
As $\vfi=2\pi m/N$ the LHS of (\ref{eq:BAE}) becomes ${\rm LHS}\rightarrow
\eE^{\mp \ir \vfi N}=1$. On the RHS the term $\mu_\infty$ drops out leaving finite
 differences $\mu_j-\mu_\kind= \ir \pi (j-\kind)/n$. Denoting $\omega=\eE^{\ir \pi/n}$, we have
\begin{align}
&{\rm RHS}_j=\prod_{\substack{\kind=1 \\\kind\neq j}}^{n} \frac{  \omega^{j-\kind}\eE^{-\ir \vfi} - \omega^{-(j-\kind)}\eE^{\ir \vfi} }
  {  \omega^{j-\kind}\eE^{\ir \vfi} - \omega^{-(j-\kind)}\eE^{-\ir \vfi} }\nonumber\\
  &=\prod_{\kind=1}^{n-1} \frac{  \omega^{\kind}\eE^{-\ir \vfi} - \omega^{-\kind}\eE^{\ir \vfi} }
{  \omega^{\kind}\eE^{\ir \vfi} - \omega^{-\kind}\eE^{-\ir \vfi} }=
\prod_{\kind=1}^{n-1} \frac{  \omega^{\kind}\eE^{-\ir \vfi} - \omega^{-\kind}\eE^{\ir \vfi} }
{-  \omega^{-\kind}\eE^{\ir \vfi} + \omega^{\kind}\eE^{-\ir \vfi} }=1. \nonumber
\end{align}
Here  we  used that the set of $\omega^{j-\kind}$ with $\kind=1,...,n$ (and $\not=j$)
is identical to the set of $\omega^{\kind}$ with $\kind=1,...,n-1$ as we have $\omega^n=-1$.
~~~

\textit{Phantom Bethe vectors for periodic chains.} The Bethe vectors corresponding to the phantom Bethe roots (PBR)
solution (\ref{Res:PhantomBetheRoots}), under the conditions of Lemma 1,
can be constructed as described below
(\ref{singlspartmoment}).
The two signs $\pm$ in (\ref{Res:PhantomBetheRoots}) correspond to different
Bethe vectors which upon normalization read
\begin{align}
&\ket{\pm,n}\!=\!\frac{1}{n! \sqrt{\binom{N}{n}} } \sum_{\kind_1,\ldots ,\kind_n=0}^{N-1}
\eE^{\pm \ir \vfi(\kind_1+\ldots +\kind_n)}
\si_{\kind_1}^{-}\ldots  \si_{\kind_n}^{-} \ket{0}\,, \no\\
&n=0,1,\ldots,N. \label{eq:PBRstates}
\end{align}
Each multiplication by a $B(\mu_j)$-operator adds a
quasiparticle with momentum $\mom(\mu_j)$ and zero energy.  Within the
standard picture \cite{FaddeevTakhtajan,SklyaninFaddeevTakhtajan}
quasi-particles obey a ``Fermi rule'': all $\mom(\mu_j)$ are usually
different. This property is violated for phantom Bethe roots $\mu_p$ for which
all $\mom(\mu_p)$ are exactly the same: either $\mom(\mu_p)=+\vfi\equiv \mom_0$ or 
$\mom(\mu_p)=-\vfi\equiv-\mom_0$ depending on the sign of the singular part in (\ref{Res:PhantomBetheRoots}).
Repeated action of $B$
generates ``phantom" Bethe states (\ref{eq:PBRstates})
with ``quantized" momenta $\pm n \vfi$ and zero energy for all magnetization
sectors $n$, yielding the degeneracy of the eigenvalue $E=0$ between different
sectors.  Note that the $E=0$ state is not a groundstate of  (\ref{eq:XXZ}), 
which is obtained by filling the Fermi sea with  quasiparticles giving negative energy contributions to (\ref{eq:BetheEigenvalue}). 
The dimension of the degenerate subspace is $\deg=2(N-1)+2=2N$
since the states $\ket{+,n},\ket{-,n}$ for $n=1,2,\ldots N-1$ are linearly
independent and for $n=0,N$ the states $\ket{+,n},\ket{-,n}$ coincide.  The
degeneracy between sectors with different magnetization leads to eigenstates
with periodic modulations in the density profile. Indeed, the SHS
(\ref{eq:SHS}) with positive chirality and $\varphi=+\vfi= {2 \pi m}/{N}\neq
\pi $ is a linear combination of phantom Bethe states $\ket{+,n}$, and SHS
(\ref{eq:SHS}) with opposite chirality $\varphi=-\vfi$ is a linear combination
of $\ket{-,n}$
\begin{align}
\ket{SHS(y_0,\pm {2 \pi m}/{N})}={\binom{N}{n}}^{1/2}\sum_{n=0}^N \eE^{y_0 n}
  \ket{\pm,n},\label{eq:SHSexpansionInBetheVectors}
\end{align}
see Supplemental \cite{SM} for the proof.  
Finally, note that the states (\ref{eq:PBRstates}) are chiral, which is evidenced
by nonzero expectation values of the magnetization current,  see Supplemental material \cite{SM}.
\begin{align}
\bra{\pm,n} j^z \ket{\pm,n}&= \pm 
\frac{8 n (N-n)}{N(N-1)}
 \sin
\vfi\,,
\label{eq:PhantomBetheRoots-SpinCurrent}
\end{align}
 reaching its maximum of order 
$|j^z|\rightarrow 2 \sin \ga$ for $n=N/2$.

\textit{Mixtures of regular and phantom excitations for the periodic $XXZ$
  model.}  
Here we  show that phantom Bethe roots can appear  alongside with
usual finite Bethe roots,  for other special values of the anisotropy.

Let us assume that within a sector of $n_0$ flipped spins, there exists a BAE
solution with $n$ phantom Bethe roots $\mu_1, ..., \mu_n$ and the remaining
$r=n_0-n$ Bethe roots are regular. We denote the regular roots as $x_1, \ldots
,x_r$ where $x_{j}= \mu_{n+j}$.  Let us consider separately the BAE
(\ref{eq:BAE}) subsets for phantom $\mu_p$ and for regular $x_j$. Substituting
(\ref{Res:PhantomBetheRoots}) in (\ref{eq:BAE}) we obtain
\begin{align}
&\eE^{ \ir \vfi (N-2 r)}= 1,\label{speclrootsII}
\end{align}
since each factor of the RHS containing a mixed pair $\mu_p$, $x_j$
contributes a term $\exp(2\ir\vfi )$. The product over factors
of the RHS involving two phantom roots results in $+1$ precisely
as in \textbf{ Lemma 1}.  The criterion
(\ref{speclrootsII}) fixes the anisotropy parameter 
while the BAE subset for regular roots simplifies to
\begin{align}
&\frac{\sinh^N (x_j\!-\!\ir \vfi/2)}{\sinh^N (x_j\!+\!\ir \vfi/2)} \!=\!
\eE^{\pm 2 \ir \vfi n} \prod_{ \substack{l=1\\\l\neq j}}^{r}
\frac{\sinh (x_j\!-\!x_l\!-\!\ir \vfi)}{\sinh (x_j\!-\!x_l\!+\!\ir \vfi)},\no
\end{align}
for all $j=1,\ldots ,r$, 
see also \cite{Ref4,Ref5}. This has the structure of the BAE of a twisted $XXZ$ chain, because of the
presence of a constant phase factor.  The signs $\pm$ match those in
(\ref{Res:PhantomBetheRoots}).

\textit{Phantom excitations in  the open $XXZ$ chain.}
The energy of 
 Hamiltonian (\ref{eq:XXZopen}) is given by  (\ref{eq:BetheEigenvalue}) with an additional offset, 
$\en=\sum_{j=1}^N e(\mu_j)+E_0$, where
\begin{align}
&E_0=-\sinh \eta
\left(\coth \al_{-}+\coth \al_{+}+\tanh \be_{-}+\tanh \be_{+}\right),\label{E-XXZopen}
\end{align}
where the  boundary fields $h_{l,r}$ are parametrized as
\begin{align}
\vec{h}= \frac{\sinh \eta}{\sinh \al_{\pm}\cosh \be_{\pm}}
(\cosh \th_{\pm}, \ir \sinh \th_{\pm},\mp \cosh \al_{\pm}\sinh \be_{\pm}),\nonumber
\end{align}
and $+ (-)$ corresponds to the right (left) field. The Bethe roots
$\mu_j$ satisfy BAE of a somewhat bulky form
\cite{OffDiagonal,Zhang2015,PhantomLong,SM}.  After some algebra
\cite{PhantomLong} we find that if
\begin{align}
\pm (\th_{+} - \th_{-}) &= (2M -N+1)\eta +\al_{-}+ \be_{-}+ \al_{+}+ \be_{+}\no\\
&  \ \mod 2\pi \ir,
\label{ConditionPhantom}
\end{align}
is satisfied with some integer $M=0,1, \ldots , N-1$, each set of $N$ Bethe
roots contains $n$ phantom Bethe roots of type (\ref{Res:PhantomBetheRoots}),
where $n$ takes one of two values $n_{+}=N-M$ and $n_{-}=M+1$
\cite{PhantomLong,ChiralBA}.  The remaining $N-n$ Bethe roots $x_j$ $(=\mu_{n+j})$ are
regular and satisfy reduced BAE
\begin{align}
&\frac{G_\pm(x_j- \frac{\eta}{2})\sinh^{2N} (x_j+\frac{\eta}{2})}{ G_\pm(-x_j-\frac{\eta}{2})\sinh^{2N} (x_j-\frac{\eta}{2})} =
\prod_{\substack{l=1\\l\neq j}}^{N-n_\pm} \frac{\sinh (x_j-x_l+\eta) }
{\sinh (x_j-x_l-\eta) }\times\nonumber\\[2pt]
&\times\frac{\sinh (x_j+x_l+ \eta)}{\sinh (x_j+x_l-\eta)},\quad j=1,\ldots,N-n_\pm,
\label{BAE1}\\
&G_\pm(u)=\prod_{\si=\pm}\sinh (u\mp\al_\si)\cosh (u\mp\beta_{\si})\,,\no
\end{align}
while the total eigenvalue  has
contributions from the regular Bethe roots only.  We like to note that
(\ref{E-XXZopen})
holds literally for case $n=n_+$. For $n=n_-$ the
$+E_0$ contribution in (\ref{E-XXZopen}) is to be replaced by $-E_0$, see
\cite{PhantomLong}. We find that the BAE (\ref{BAE1}) for $n=N-M$ describes $\dim
G_M^{+} = \sum_{m=0}^M \binom{N}{m}$ Bethe states, while the remaining $2^N -
\dim G_M^{+}$ eigenstates are contained in the other, complementary BAE set
for $n=M+1$ \cite{PhantomLong, ChiralBA}.  Unlike in the periodic setup, where some
  Bethe eigenstates contain PBR modes, and other eigenstates are fully
  regular, in open systems, satisfying criterion (\ref{ConditionPhantom}),
  \textit{all} $2^N$ eigenstates include phantom Bethe roots.  
Remarkably,  the  condition (\ref{ConditionPhantom}) appears in \cite{OffDiagonal03,Nepomechie2003,Rafael2003,Cao2013off} as a 
condition for the
application of the Algebraic Bethe Ansatz. The
BAE set (\ref{BAE1}) coincides with that found by an alternative method
\cite{OffDiagonal03,Nepomechie2003,Cao2013off}.

Now we focus on the simplest Bethe states, corresponding to all Bethe roots
being phantom, $n_{+}=N$, the respective energy given by $E_0$ 
(\ref{E-XXZopen}).  We demonstrate that such ``phantom'' Bethe states are
spin-helix states (\ref{eq:SHS}) with appropriately chosen parameters. The
phantom Bethe states for mixtures of phantom and regular Bethe roots can be
also obtained explicitly and show chiral features \cite{PhantomLong,ChiralBA}.

\textit{Phantom Bethe states: open $XXZ$ chain. Easy plane regime $|\De|<1$.}
It is straightforward to verify that the SHS (\ref{eq:SHS}), with
$\varphi=\gamma$, ${\rm Re}[y_0]=\be_{-}$ and phase ${\rm Im}[y_0]=\pi + \ir
\al_{-}+ \ir \th_{-}$ (note that $\al_{-},\th_{-}$ are imaginary and
$\be_{-}=-\be_{+}$ are real to ensure hermiticity of $H$), is an eigenstate of
$H$.  Indeed, one can check that (\ref{eq:hlAction}), (\ref{eq:hrAction})
are satisfied with $\la_{\pm}=-\sinh\eta( \coth \al_{\pm} - \tanh \be_{\pm})$,
so that this SHS is an eigenvector of (\ref{eq:XXZopen}) with eigenvalue
$\la_{-} +\la_{+} $, which coincides with the phantom Bethe vector eigenvalue
$E_0$  (\ref{E-XXZopen}). For the magnetization profile of this SHS
see Fig.~\ref{Fig-SHSCosCosh}, top panel.  Unlike for the periodic chain,
here the eigenvalue $E_0$ is generically non-degenerate.

\textbf{Simplest experimental setup.}  Using our results, long-lived SHS can
be obtained in experiments where effectively one-dimensional spin $1/2$ $XXZ$
chains with tunable anisotropy are realized
\cite{2020NatureSpinHelix,2021Ketterle}. A spin helix of the form
  (\ref{eq:SHS}) with an adjustable wavelength is created within cold atoms
  setups by applying a magnetic field gradient in $z$ direction on an array of
  initially noninteracting qubits polarized along the $x$ axis, see Methods of
  \cite{2020NatureSpinHelix} for details.  To make the SHS an eigenstate of
  the $XXZ$ Hamiltonian, the wavelength $Q$ of the spin-helix and the
$z$-anisotropy $\De$ must be related as $\De=\cos Q a$ where $a$ is the
lattice constant.  Indeed under this choice an SHS of type (\ref{eq:SHS})
$\ket{SHS_{\pm}} := \ket{SHS(\ir F_0, \pm Q a)}$ will remain invariant in the
bulk and change initially only at the boundaries, since
\begin{align}
&\sum_{n=1}^{N-1}  h_{n,n+1}(\De)  \ket{SHS_{\pm}}  = 
\mp \ir \sin Q a \left(\si_1^z - \si_N^z \right) \ket{SHS_{\pm}},\nonumber
\end{align}
as follows from (\ref{eq:Divergence}). The ends of the spin chain will thus
play the role of defects, and the state in the bulk will be altered only by
propagation of the information from the boundaries.  Thus the state can be
 destroyed only after times of order $t = N a/v_{char}$, where $v_{char}$
is the sound velocity, $N$ is the  number of spins and $a$
is the lattice constant.  For example, in 
\cite{2020NatureSpinHelix,2021Ketterle}, the process of the expansion of the defect in the
bulk can be monitored. On the other hand, if the SHS period does not match the
anisotropy $\De\neq\cos Q a$, then the initial SHS will be
destroyed after times of order $t = a/v_{char}$.  On one hand, the effect
is robust (w.r.t.~the phase of the helix  and chain length $N$), and on
the other hand, it is sensitive w.r.t.~the matching condition for the
anisotropy $\De$.  This sensitivity can be used as a benchmark for calibrating
the anisotropy, or the wave-length of the produced SHS, or both.

~~~
\textit{Phantom Bethe states:  Easy axis  $\De=\cosh \eta>1$.}
The SHS of the form (\ref{eq:SHS}) with 
$y_0= \ir \pi - \th_{-}+  \al_{-}+ \be_{-}$ 
satisfies (\ref{eq:hlAction}), (\ref{eq:hrAction}) with $\ka \rightarrow
-\sinh \eta$ and $\la_\pm=- \sinh\eta ( \coth \al_{\pm} + \tanh
\be_{\pm})$. Consequently, state (\ref{eq:SHS}) is an eigenstate of $H$ with
eigenvalue $\la_{+}+ \la_{-}= E_0$.  Thus, state (\ref{eq:SHS}) is a
phantom Bethe vector.  It describes spins on the lattice with fixed azimuthal
angle and changing polar angle along the chain, see 
Fig.~\ref{Fig-SHSCosCosh}, lower panel.  Unlike the ``azimuthal'' spin helix
state (\ref{eq:SHS}), the ``polar'' SHS  carries no spin
current, $\langle j^z\rangle_{SHS_{polar}}=0$.

\textit{Discussion.} We have described a novel scenario of excitations in
integrable systems, namely phantom  excitations with
phantom Bethe roots  corresponding to unbounded rapidities.  The existence
criterion for these states is formulated and depends on the boundary
conditions of the system. Under this criterion a certain subset of
Bethe roots is located at infinity with relative positions at equidistant
points. This resembles a perfect TBA string, but is of entirely different
nature.  

For models with periodic boundaries the PBR are responsible for degeneracies
between sectors with different total magnetization, and lead to factorized
spin helix eigenstates at anisotropies given by (\ref{speclrootsII}). 
Also
for the open $XXZ$ model the PBR related eigenstates are
spin helix states with winding polarization vector, in the easy plane regime,
and the ``polar angle''-version of the latter, in the easy axis regime.  Our
results can be used for the generation of stable spin helix states in
experimental setups realizing $XXZ$ chains
\cite{2020NatureSpinHelix,2021Ketterle}.

While our discussion was restricted to the $XXZ$ model, the presence of
phantom Bethe roots, due to their simple analytic form
(\ref{Res:PhantomBetheRoots}), can be easily established in other integrable
models, e.g. in the periodic spin-1 Fateev-Zamolodchikov model
\cite{ZF80,Kulish83,Ritt90}, and arbitrary spin $s$ generalizations
\cite{,ZF-BAE,Babujian1986,Kirillov1987,FaddeevArXiv}, see Supplemental Material \cite{SM}.  It
would be interesting to search for PBR analogues in intrinsically
non-hermitian integrable models, e.g.  \cite{EsslerZiolkowska,Pozsgay}.


\begin{acknowledgments}
 Financial support from the Deutsche Forschungsgemeinschaft through DFG
 project KL 645/20-1, is gratefully acknowledged.  X.~Z. thanks the Alexander
 von Humboldt Foundation for financial support. V.~P. acknowledges support by
 European Research Council (ERC) through the advanced Grant
 No. 694544—OMNES. V.~P. thanks S.~Essink for indicating the work in
 \cite{2020NatureSpinHelix}. We thank W. Ketterle for drawing our
   attention to his newest experiment \cite{2021Ketterle}, where the time
   evolution of a transversal spin helix state (\ref{eq:SHS}) has been
   studied.
\end{acknowledgments}


\end{document}